\documentclass[twocolumn,showpacs,prb,floatfix,superscriptaddress,citeautoscript
,cite]{revtex4}
\usepackage{graphicx}
\usepackage{color}
\usepackage{amsmath}

\usepackage{booktabs}

\begin{document}

\title{Electronic, Optical and Mechanical Properties of Silicene Derivatives}

\author{M. Yagmurcukardes}
\email{mehmetyagmurcukardes@iyte.edu.tr}
\affiliation{Department of Physics, Izmir Institute of Technology, 35430 Izmir,
Turkey}

\author{C. Bacaksiz}
\affiliation{Department of Physics, Izmir Institute of Technology, 35430 Izmir, 
Turkey}

\author{F. Iyikanat}
\affiliation{Department of Physics, Izmir Institute of Technology, 35430 Izmir, 
Turkey}

\author{E. Torun}
\affiliation{Department of Physics, University of Antwerp, Groenenborgerlaan 
171, B-2020 Antwerp, Belgium}

\author{R. T. Senger}
\email{tugrulsenger@iyte.edu.tr }
\affiliation{Department of Physics, Izmir Institute of Technology, 35430 Izmir, 
Turkey}

\author{F. M. Peeters}
\affiliation{Department of Physics, University of Antwerp, Groenenborgerlaan
171, B-2020 Antwerp, Belgium}

\author{H. Sahin}
\email{hasansahin.edu@gmail.com}
\affiliation{Department of Physics, University of Antwerp, Groenenborgerlaan 
171, B-2020 Antwerp, Belgium}
\affiliation{Department of Photonics, Izmir Institute of Technology, 35430 
Izmir, 
Turkey}
\date{\today}

\begin{abstract}
Successful isolation of graphene from graphite opened
a new era for material science and condensed matter physics. Due to this 
remarkable achievement, there has been an immense 
interest to synthesize new two dimensional materials and to investigate 
their novel physical properties. Silicene, form of Si atoms arranged in a 
buckled honeycomb geometry,
has been successfully synthesized and emerged as a promising material 
for nanoscale 
device applications. 
However, the major obstacle for using silicene in electronic 
applications is the lack of a band gap similar to the case of graphene. 
Therefore, 
tuning the electronic properties of silicene by using chemical 
functionalization methods such as hydrogenation, halogenation or oxidation
has been a focus of interest in silicene research. In this paper, we 
review the recent studies on the structural, electronic, optical 
and 
mechanical properties of
 silicene-derivative structures. Since these derivatives have various band gap 
energies, they are promising candidates for the next generation of electronic 
and 
optoelectronic device applications.
\end{abstract}

\maketitle

\section{Introduction}

Layered bulk materials consisting of two dimensional (2D) sheets which are
hold together with weak, interlayer van der Waals interaction have been the 
focus
of interest for more than a century\cite{Brodie,Peierls1,Peierls2}. With the 
advancement of synthesis and characterization techniques it has been possible 
to isolate ultra thin films down to a monolayer of these materials 
which 
became feasible in the last decade. 
Monolayer forms of these layered bulk materials often exhibit different 
physical properties than their bulk counterparts.
The first isolated 2D material is known to be graphene, a one-atom-thick carbon 
sheet, with
extraordinary physical properties\cite{Novo1,Novo2,Geim1}. After the successful
exfoliation of graphene by Novoselov and Geim, researchers have been 
searching for several other 2D materials that can exist in single layer form 
such as hexagonal 
monolayer crystals III-V binary 
compounds\cite{hasan1,Golberg,Zeng,Song,Bacaksiz,Zhuang,QWang,KKim,MFarahani},
transition metal dichalcogenides (TMDs)\cite{Wang,Wilson,Horzum,Bacaksiz2} and 
the group IV
elements (silicene, germanene, 
stanene)\cite{Cahangirov,Vogt,Lin,Fleurence,Davila,Zhu}. Among these 2D
monolayer materials, graphene and silicene are known to posses semi-metallic
character while the members of TMDs family compounds generally display 
semiconducting
behavior with a band gap of 1-2 eV. In all of these 2D materials silicene 
occupies an important position for the next generation of nanoscale 
technology which up to now is mostly based on silicon.

According to its electronic-band structure, graphene has a semi-metallic 
character 
which is not suitable for optoelectronic applications. One possible way to 
open a gap in the band structure of graphene is to functionalize its surface 
with various types of 
atoms such as H, F and Cl which were widely studied and successfully 
synthesized. It was shown that both full and partial hydrogenation of 
graphene
leads to semiconducting materials with different 
band gap values\cite{Boukhvalov,Haberer,Sofo}. Similar to the hydrogenation 
case, experimental and
theoretical studies showed that the band gap of fluorinated-graphene can 
alter from 0 to 3 eV depending on the fluorination level. 
\cite{cRobinson,cSamarakoon,cCheng1,cJeon,cGarcia}

Silicene, a 2D honeycomb structure of Si atoms with a buckled geometry, has been
attracting great interest due to its physical properties such as 
possessing massless Dirac fermions and large spin-orbit coupling resulting in 
an 
intrinsic band gap\cite{Kara,Xu1}. 
The buckled structure of silicene is a 
consequence of sp$^2$-sp$^3$ hybridization of Si atoms. This makes the 
structure 
of silicene different 
from the flat structure of graphene. 
Another important physical property of silicene is its high surface reactivity 
which widens the methods of 
manipulating its electronic, magnetic and mechanical properties\cite{hasan2}. 
Thus, the 
functionalization of
its surface and applying external mechanical strain are some of the widely used 
ways of
controlling the electronic properties of silicene for its practical usage in 
device
technology.

After the theoretical prediction and successful
synthesis of silicene, researchers have focused on 
doping\cite{Lin2,Quhe,Ni,Cheng,Sivek,Zheng},
chemical 
modification\cite{hasan2,Okamoto1,Nakano,Okamoto2,Sugiyama,Pereda,Spencer,Du}
and strain 
engineering\cite{Liu12,Qin12,Zhao12,Hu13,Kal-13,Dur14,Moh14,Hus14,Wang14,Zhu14,
Yang14,Cao15}
 in order to modify its electronic structure. Studies have
demonstrated that fully hydrogenated silicene is a 
semiconductor\cite{Ding,Voon,Houssa} while half hydrogenated
silicene is still a semi-metal or direct-gap semiconductor depending on the 
hydrogenation configuration\cite{Zhang}. Functionalization of silicene with 
halogen atoms (F, Cl, Br and I) was also considered in several studies for
tuning its electronic structure\cite{Wei,Gao,Zhang2,Wang2}. Studies on
fully halogenated silicene indicated that it
possesses a direct-gap semiconducting character with various band gap depending 
on the type of the halogen atom. Other functionalization methods like 
doping organic molecules on hydrogenated silicene have also been considered. 
\cite{Okamoto1,Nakano,Okamoto2,Sugiyama} 
Moreover, the oxidation of silicene was studied both theoretically and
experimentally by several research groups which is important for the use of
2D materials in nanoscale device technology.  
\cite{Padova1,Padova2,Molle,Friedlein,Xu,Liu}

In this review we summarize the studies on the structural, electronic, optical
and mechanical properties of silicene derivatives. This review is organized 
as follows:
We first provide the physical properties of hydrogenated
silicene in Sec. \ref{H}, the oxidized silicene in Sec. \ref{Ox} and  the 
halogenated silicene in Sec \ref{Halo}.
The physical properties of silicene functionalized with organic molecules are 
given in Sec. \ref{Organic} while the properties of silicene decorated with 
adatoms are given in Sec. \ref{doping}. Finally we present a brief summary in Sec. \ref{summ}

\begin{figure}[htbp]
\includegraphics[width=8cm]{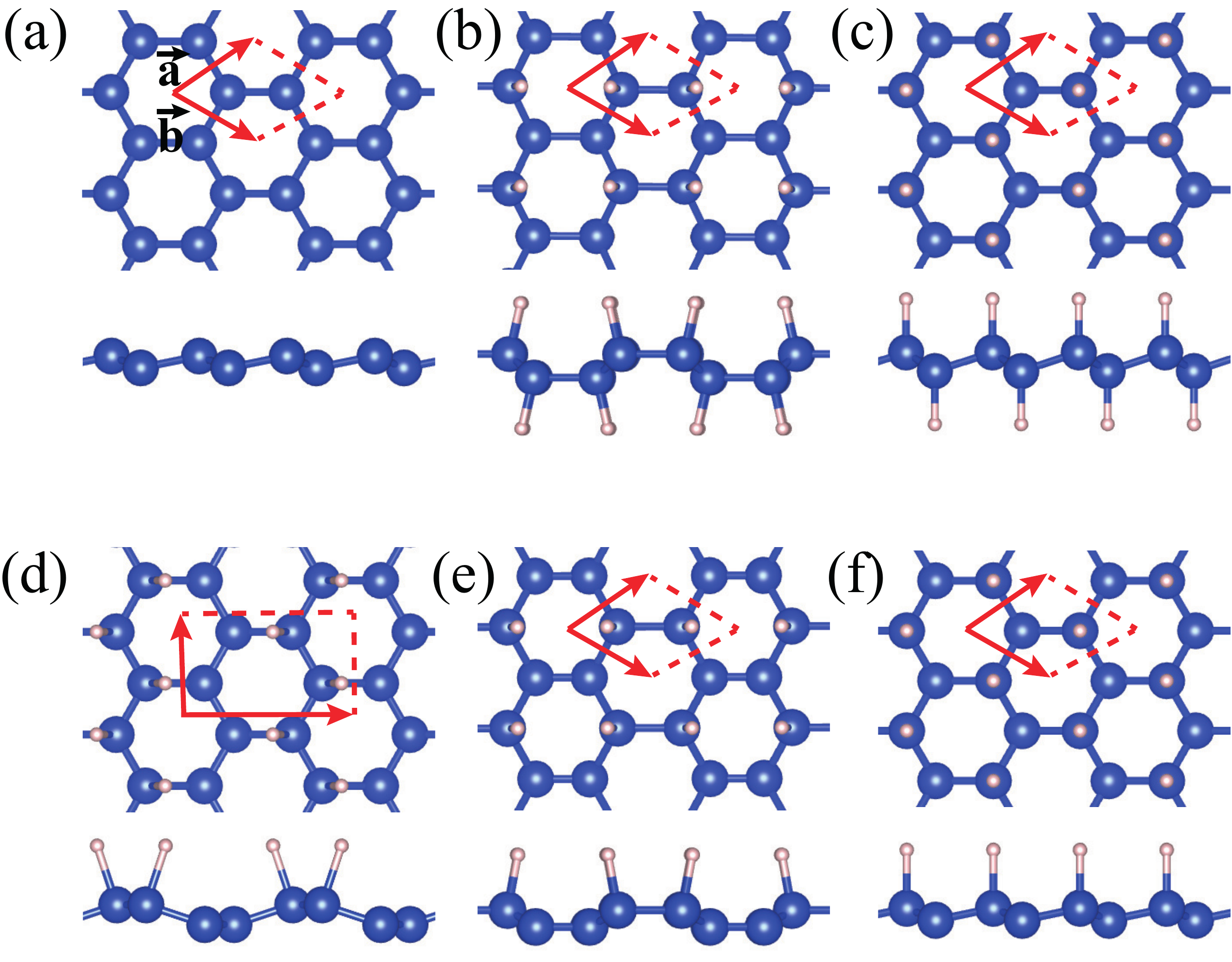}
\caption{\label{H-stuc}
(Color online) Top and side views of optimized structures of (a) silicene;
(b)-(c) boat-like and chair-like full-hydrogenated silicene; (d)-(f) zigzag,
boat-like and chair-like half-hydrogenated silicene, respectively. The 
primitive unit cell of each structure
is shown by red lines. Si and H atoms are shown in blue and grey, respectively.}
\end{figure}


\section{HYDROGENATED SILICENE}\label{H}

In this section we review the results of studies on hydrogenation of
silicene. Like C atoms in graphene, Si atoms in silicene have unpaired
electrons which are suitable for possible functionalizations. Among these
possible functionalizations, hydrogenation was studied
extensively in the literature. \cite{Zhang,Drissi1,Drissi2,Zhang3} It has been 
shown 
that 
two possible configurations exist for the hydrogenation
process of silicene, fully-hydrogenation (fH), namely silicane, and single side 
hydrogenation, half-hydrogenation (hH), 
similar to the case of graphene. 

\begin{figure*}[htbp]
\includegraphics[width=16.0cm]{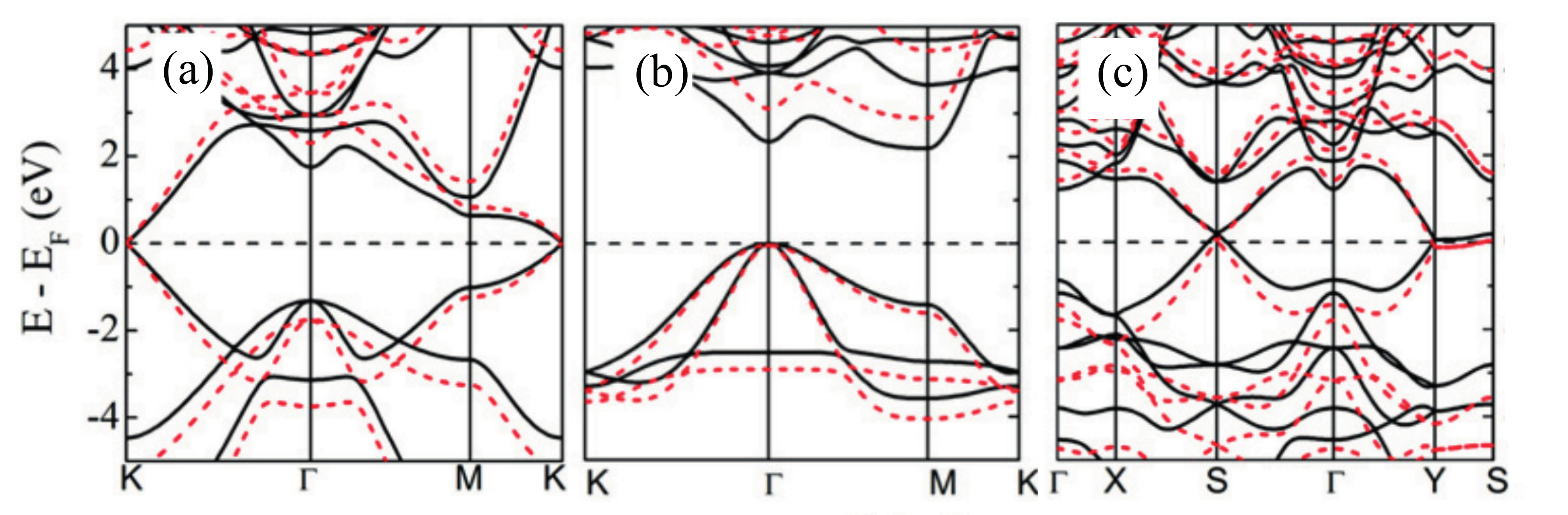}
\caption{\label{H-band}
(Color online) Energy-band structures of (a) bare silicene, (b) chair-like
silicane, and (c) zigzag HH silicene (taken from Ref. \cite{Zhang})}
\end{figure*}

In Fig. \ref{H-stuc}, possible geometric structures are given for bare,
fully and half hydrogenated silicene crystals. Zhang \textit{et al.}
investigated the structural properties of fH and hH cases of silicene
by first principles calculations\cite{Zhang} and found that for the silicane 
structure the chair-like configuration (see Fig.
\ref{H-stuc}(c)) is the ground state and it has 30
meV/atom lower energy than the boat-like one (see Fig. \ref{H-stuc}(b)) as
confirmed by total energy calculations. The Si-H bond length was calculated to
be 1.50 \AA {} for the chair-like structure. For the hH silicene they
reported that the
zigzag structure (Fig. \ref{H-stuc}(d)) is the most stable configuration
with a total energy of 33 meV/atom and 180 meV/atom lower than
the boat-like and chair-like structures, respectively. In addition, Osborn
\textit{et al.} reported that the fH silicene structure has a higher
buckling than its bare form\cite{Osborn}. They calculated the buckling height
of fH silicene to be 0.74 \AA {} while 0.54 \AA {} was reported 
for the bare silicene case. This structural change occurs due to
the interaction between Si and H atoms which
widens the structure in the vertical direction. The buckling height of the hH 
silicene
structure is reported to be less than that of silicane
as expected\cite{Zhang4}.

Hydrogenation plays an important role for tuning 
the electronic structure of a 2D material. For instance in contrast to bare 
graphene, hydrogenated graphene, namely graphane, is a 
semiconductor\cite{Sahin3}. The same functionalization process was studied in 
the case of silicene. 
Zhang \textit{et al.} reported that
the electronic-band structure of silicene can be tuned through 
hydrogenation. It was found that silicane is an indirect-gap
semiconductor with its valence-band maximum (VBM) and conduction band
minimum (CBM) residing at the $\Gamma$ and M points, respectively. 
The band gap of silicene was found to be 2.36 eV 
within GGA approximation while it is reported as 3.51 eV by using HSE06 
functional.
These results were also predicted and supported by many other
studies\cite{Ding,Houssa,Zhang,Osborn}. In contrast to silicane,
the HH silicene crystal possesses metallic character in its zigzag structure
(see Fig. \ref{H-band}(c)).
However, the other two configurations, boat-like and chair-like structures,
were reported to be direct-gap semiconductors\cite{Zhang}.

\begin{figure}[htbp]
\includegraphics[width=6.5cm]{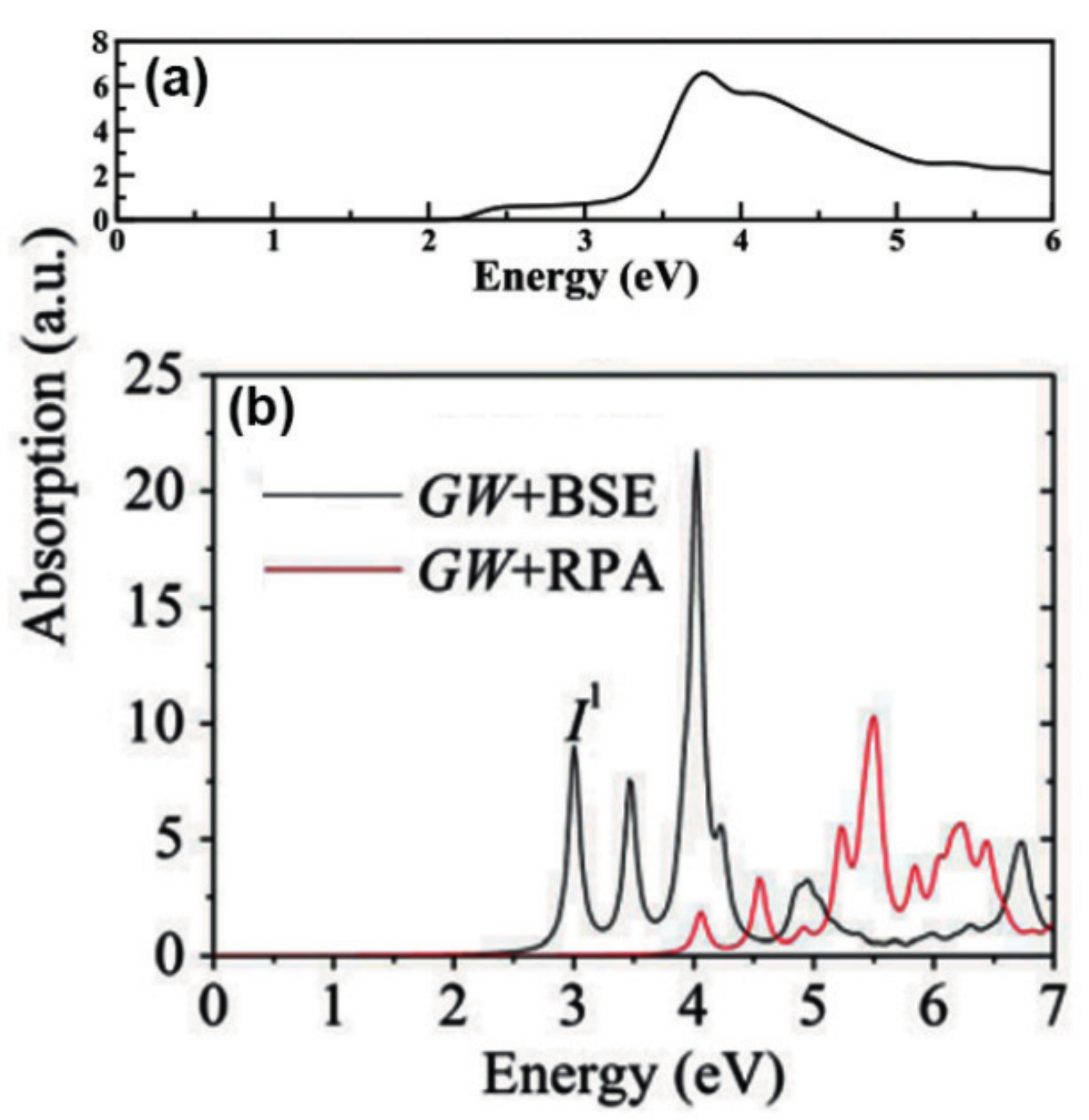}
\caption{\label{opH}
(Color online) Optical absorption spectra of silicane
calculated with (a) GGA approach \cite{Chinnathambi} and (b) GW 
approximations with (GW+BSE) or
without (GW+RPA) considering excitonic effects (taken from Ref. \cite{Rong})} 
\end{figure}
 
Optical properties of silicane such as its optical
absorption spectrum and dielectric function were 
investigated in the literature before\cite{Wei,Chinnathambi}. In addition,
the optical properties of bilayer and few layer
fH silicene structures were also predicted\cite{Huang,Liu2}. Chinnathambi
\textit{et al.} studied the optical properties of silicane
by calculating the optical absorption spectrum\cite{Chinnathambi}. They reported
that a transition from semi-metallic to semiconducting behavior is seen. The 
reason is the broken $\pi$ bonds
in silicene due to the saturation by H atoms. As seen in Fig. \ref{opH}(a), an
absorption onset at 2.2 eV was predicted which is consistent with the bandgap 
of silicane as 
calculated within the GGA approximation\cite{Chinnathambi}. Moreover, Wei
\textit{et al.} investigated the optical absorption spectra of
silicane by GW
approximation with random phase approximation (GW+RPA) and Bethe-Salpeter
equation (GW+BSE)\cite{Wei}. It was reported that the hydrogenation process 
removes conduction
at the Dirac point and causes a finite band gap opening. 
It was also reported that the GW+RPA and GW+BSE methods give different 
absorption spectra 
due to the large self-energy correlations of electrons (see Fig.
\ref{opH}(b)). The absorption onset obtained with
GW+RPA is located at about 4 eV consistent with the band gap value
calculated within HSE06 functional\cite{Houssa,Zhang}. Including the excitonic
correlations of electrons and holes, within the GW+BSE approximation,
the excitonic effect significantly shifts the onset of the absorption spectrum 
towards lower energy
(see Fig. \ref{opH}(b)).

\begin{figure}[htbp]
\includegraphics[width=8cm]{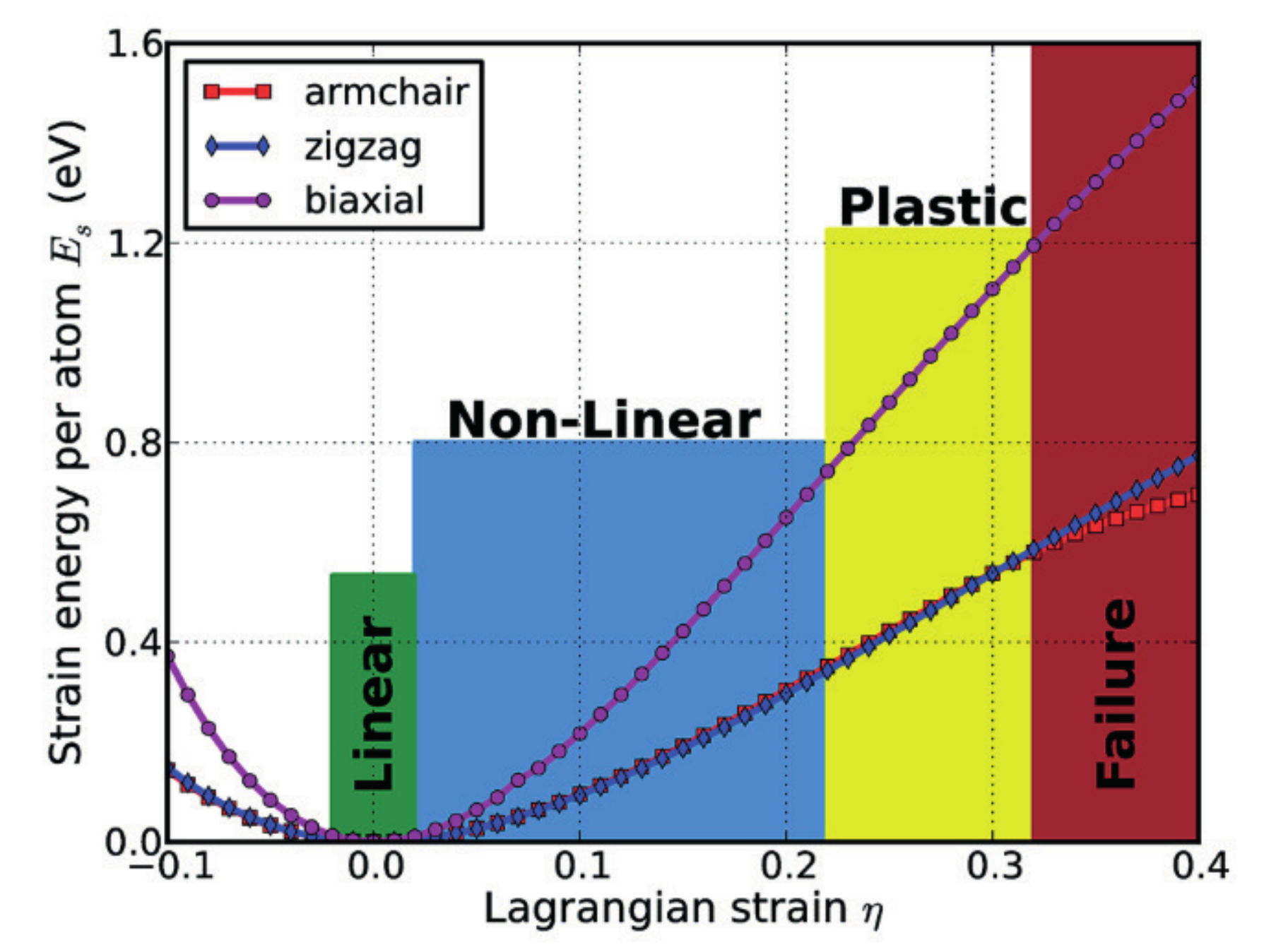}
\caption{\label{mechH}
(Color online) The strain energy per Si-H pair as a
function of applied uniaxial Lagrangian strain $\eta$ along armchair and zigzag 
directions, and biaxial
Lagrangian strain along both directions (taken from Ref. \cite{Peng})}
\end{figure}

In addition to the electronic and optical properties, the mechanical properties
of silicane were also investigated theoretically in the previous
studies\cite{Jamdagni,Peng,Yang}. Peng \textit{et al.} reported that the
in-plane stiffness (58 N/m) and Poisson ratio (0.24)
values for silicane are reduced by 16\% and 26\%, respectively, when
compared to those of silicene\cite{Peng}. The elastic limits in terms of
ultimate tensile strains were found to be 0.22, 0.28, and 0.25
along armchair, zigzag, and biaxial directions, respectively. It was reported
that these values increases by 9\%, 33\%, and 24\%,
respectively from silicene to silicane. Moreover, Jamdagni \textit{et
al.} reported that the band gap of silicane reduces
to zero with increasing applied biaxial tensile strain leads to a
semiconducting to metallic transition for silicane. Their
calculations indicated that at 2\% of tensile strain, the magnitude of the 
bandgap
first
increases to 2.22 eV and the indirect band-gap character of silicane changes to 
a
direct bandgap. Then with every 2\% increment of
tensile strain, the band gap decreases nearly by 0.3 eV and the 22\% value of 
the
strain is the critical value for semiconducting-to-metallic 
transition\cite{Jamdagni}.

\begin{figure}[htbp]
\includegraphics[width=8cm]{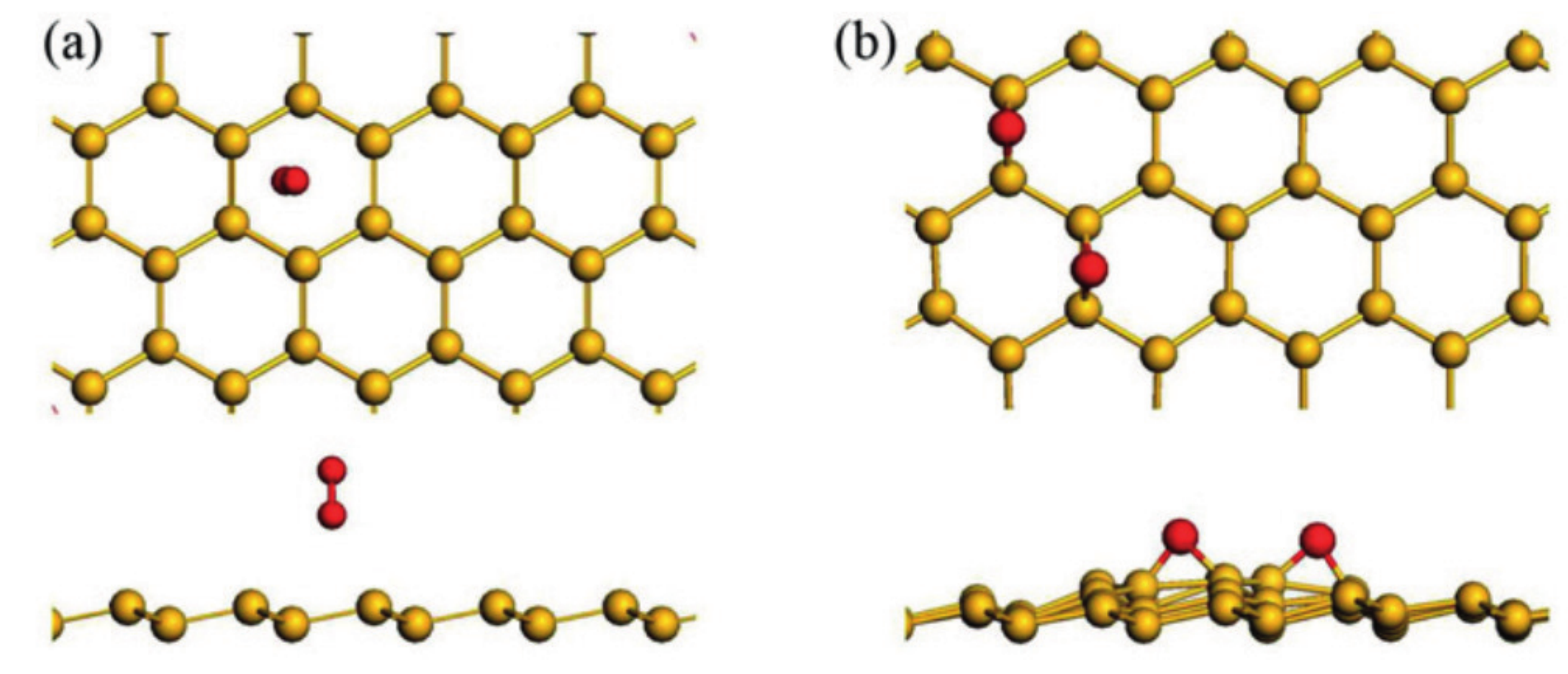}
\caption{\label{O-stuc}
(Color online) Geometric structure of oxygen molecule adsorption and 
dissociation on
 pristine silicene (a) before and (b) after relaxation (taken from Ref. 
\cite{Liu})}
\end{figure}

\section{OXIDATION OF SILICENE}\label{Ox}

Oxidation has important consequences on the usage of materials in real life 
device technologies. Thus, the oxidation processes of both bulk and 2D materials
were widely studied and investigated in previous works. Similar to all
materials, the oxidation of silicene is an important question for scientists 
during the 
fabrication of silicene-based devices.
Therefore, the possibilities of silicene oxide formation and the effects of 
oxidation on the physical properties of
silicene were studied both experimentally\cite{Du,Padova2,Molle,Friedlein,Xu}
and theoretically\cite{Padova1,Liu,Wang3,Ongun1,Ongun2,Gurel}.

\begin{figure}[htbp]
\includegraphics[width=8cm]{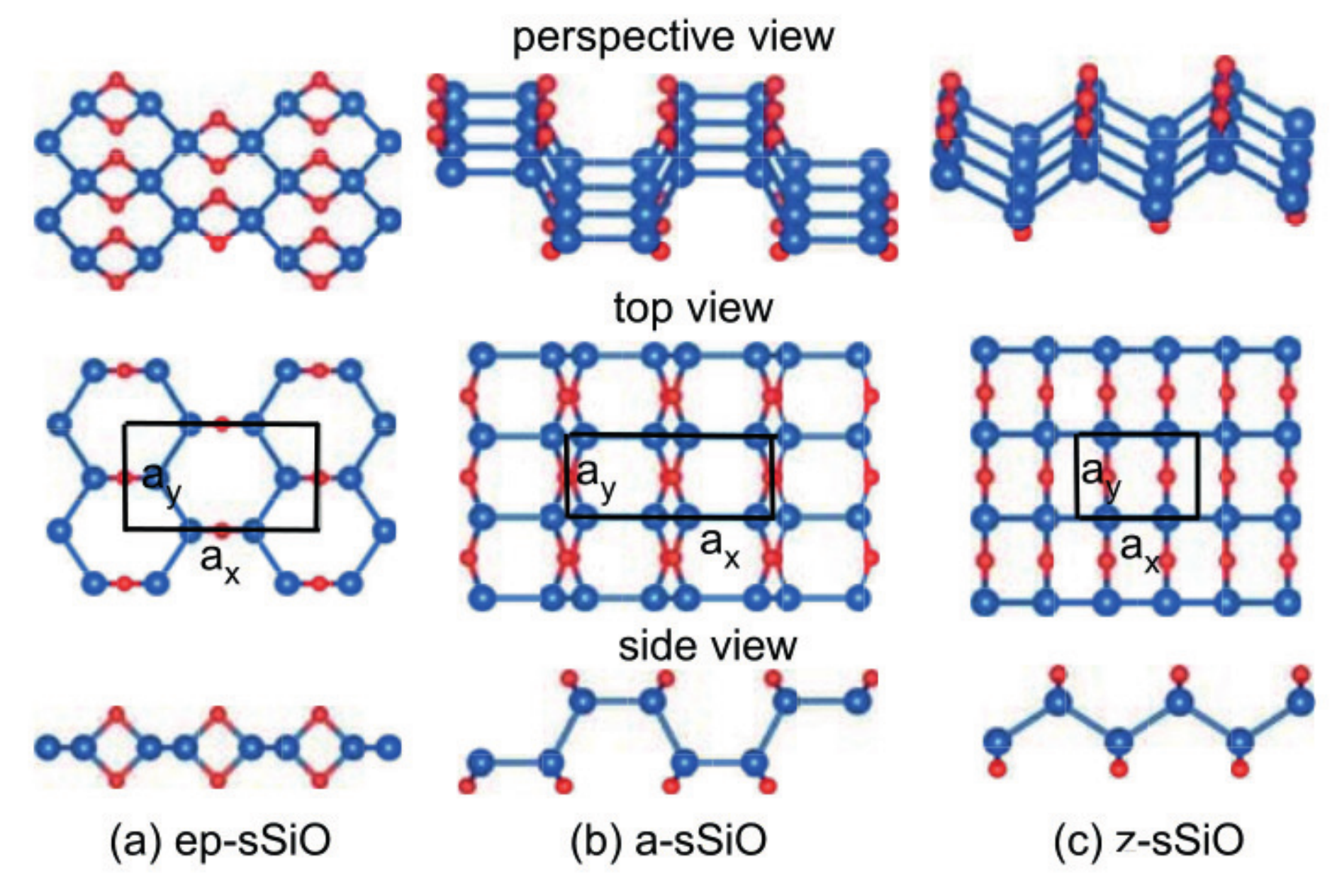}
\caption{\label{O-stuc2}
(Color online) Atomic structures of sSiO with (a) ep-conformation (O atoms
are located as
epoxy-pair groups), (b) a-conformation, and (c)
z-conformation (taken from Ref. \cite{Wang3}).}
\end{figure}

\begin{figure}[htbp]
\includegraphics[width=6.5cm]{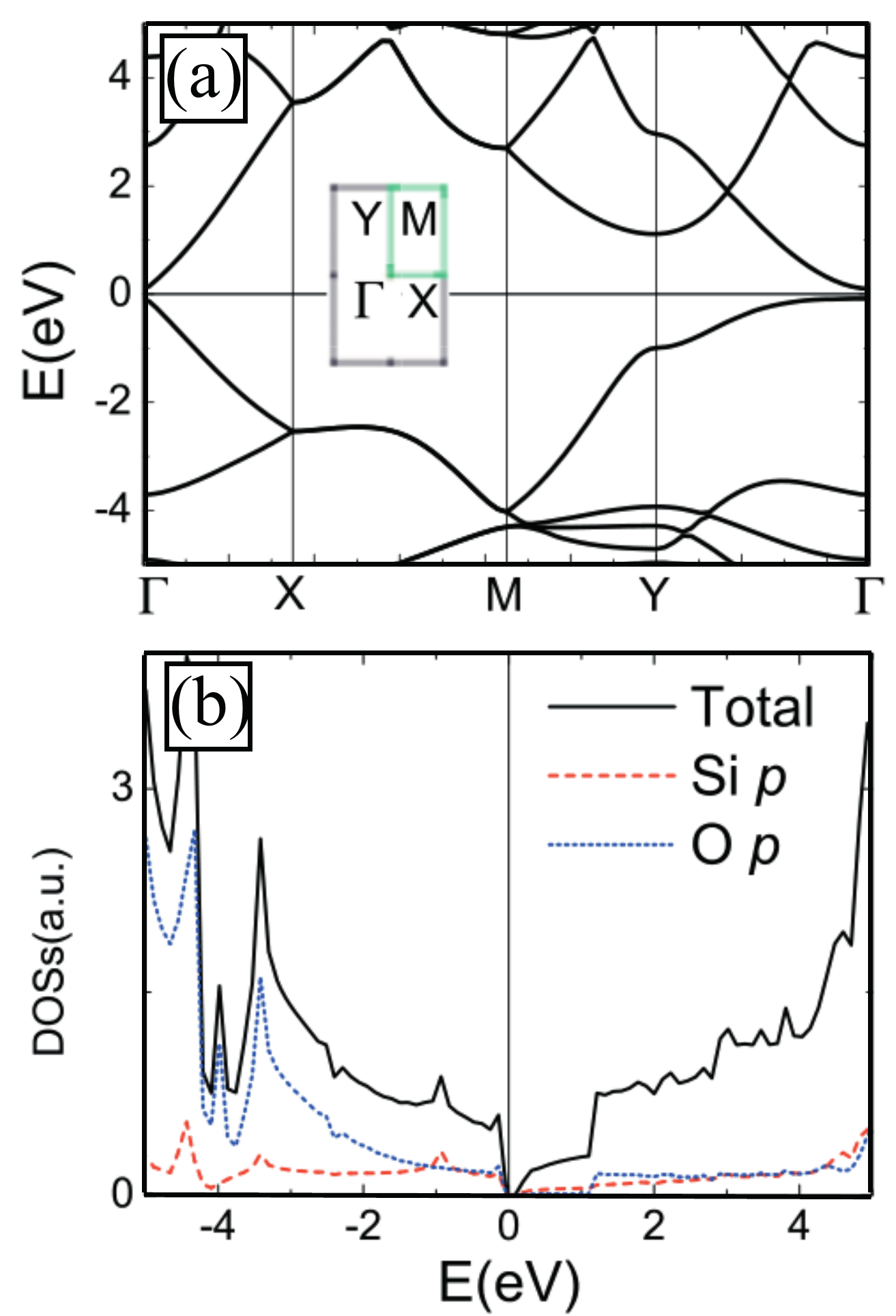}
\caption{\label{O-band}
(Color online) (a) Energy-band structure of z-sSiO calculated within GGA and
(b) the corresponding partial density of states (taken from Ref. \cite{Wang3})}
\end{figure}

Liu \textit{et al.} investigated the
oxygen adsorption and dissociation on a free-standing silicene
monolayer\cite{Liu}. It was reported that the O$_2$ molecule dissociates into O 
atoms 
on free-standing silicene and the formation of Si-O compound occurs. Also 
it was pointed out that
the oxidation of silicene is easy because of the very low energy barrier for the
O$_2$ molecule to dissociate into O atoms. Depending on the initial vertical
distance of the O$_2$ molecule to the silicene layer, the resultant O atoms can
bind to different sites of silicene. Among these sites the lowest
energy configuration is the one for which the two O atoms reside on
bridge sites of silicene (see Fig. \ref{O-stuc}(b)). The dissociation of O$_2$ 
molecule on free-standing silicene 
is confirmed by Ozcelik \textit{et al.}\cite{Ongun1} The 
Si-O bond lengths were calculated to be 1.71 \AA {} and 1.73 \AA {} for upper
and lower Si atoms, respectively. The possible migration
paths for an O atom from one bridge site to a neighboring bridge site 
exhibit energy barriers of 1.05 eV and 1.18 eV 
energy barriers. These are large values as compared to those for a graphene 
surface\cite{Liu}. In another study, the single layer
phase of silica, SiO$_2$, was predicted as a stable honeycomblike structure by
Ozcelik \textit{et al.}\cite{Ongun2}

\begin{figure*}[htbp]
\includegraphics[width=14cm]{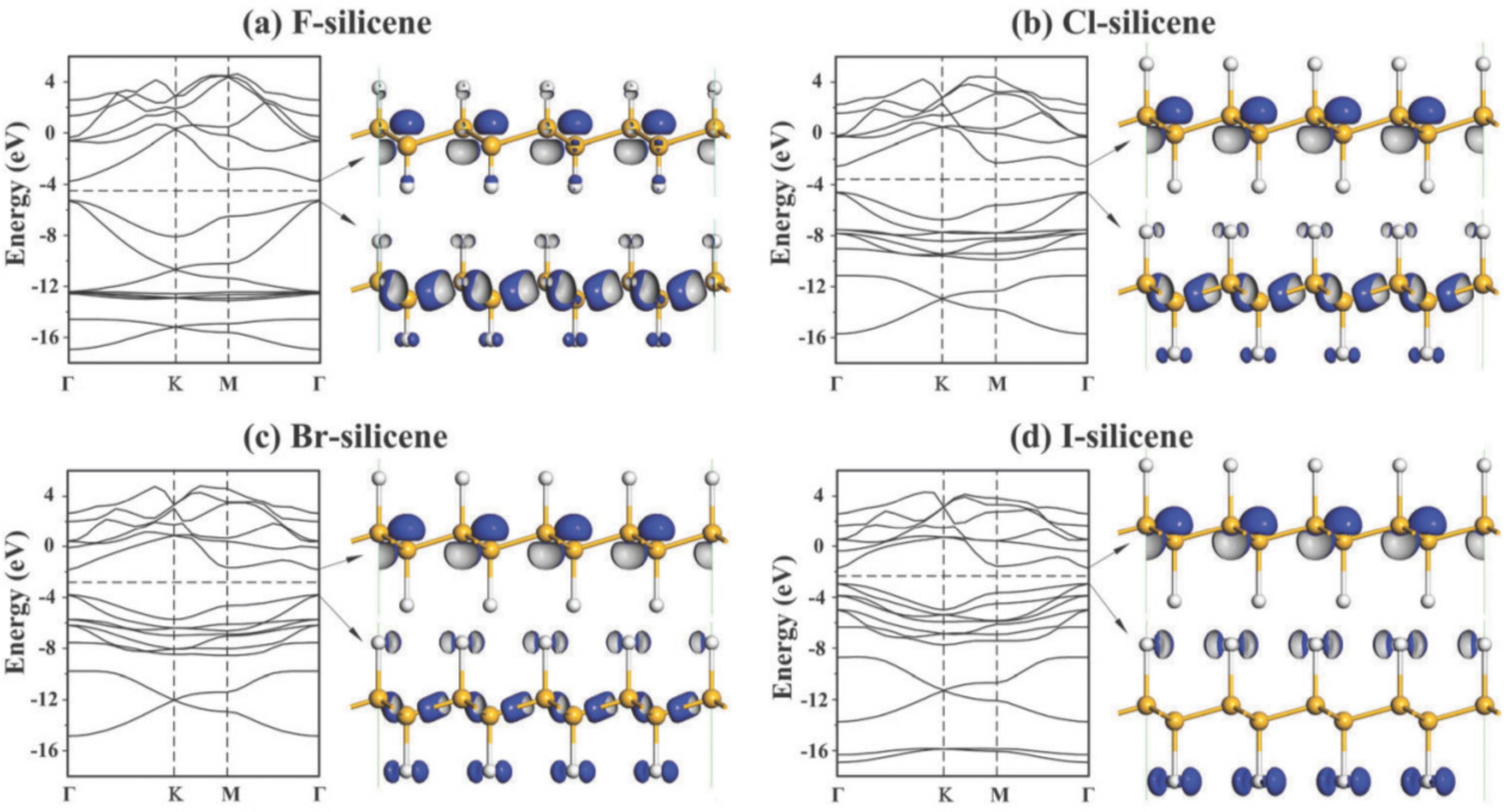}
\caption{\label{halog_band}
(Color online) The electronic band structures (at the left panel) and charge 
density
distributions (at the right panel) of the CBM (upper) and the VBM (lower) 
states at the 
$\Gamma$ point
for (a) F-silicene, (b) Cl-silicene, (c) Br-silicene, and (d) I-silicene 
structures. Dashed line indicates the Fermi level and the 
isosurface value is 0.025 
au (arbitrary 
unit) (taken from Ref. \cite{Gao})}
\end{figure*}

The effect of oxidation on the electronic properties of silicene was
investigated by Wang \textit{et al.}\cite{Wang3} They studied fully
oxidized silicene with stoichiometric ratio of Si:O = 1:1. The zigzag ether-like
conformation stoichiometric silicene oxide (z-sSiO) was found to be the ground
state configuration (see  Fig. \ref{O-stuc2}(c)). The z-sSiO configuration has 
14 and 165 meV/atom lower energy than the a-conformation and the
ep-conformation, respectively. They reported that the z-sSiO structure is a
semiconductor with a direct band gap of 0.18 eV as calculated within the GGA
approximation (see Fig. \ref{O-band}(a)) while it is found to be 1.05 eV when 
the
HSE06 functional is considered. In addition, Ozcelik \textit{et
al.} reported that single O adsorption on a silicene layer results in a
direct-gap semiconducting structure with a band gap of 0.21 eV\cite{Ongun1}. In
the study of Ozcelik \textit{et al.}, the new phase of SiO$_2$, monolayer
silica, was found to be a direct-gap semiconductor with a relatively large band
gap of 3.3 eV when compared to O-doped silicene layer\cite{Ongun2}.

Wang \textit{et al.} studied the mechanical properties of the
stoichiometric SiO structure. They found that the z-sSiO monolayer has some
prominent elastic characteristics, as negative Poisson ratios and 
exhibits an unconventional auxetic
behavior. When these auxetic materials are stretched in one direction, they 
become thicker in the perpendicular direction.
The reason for this auxetic behavior is the assembly of Si-O bonds into bending 
-O-
network along the y-direction. The mechanical properties of monolayer silica 
were
investigated in terms of in-plane stiffness and Poisson ratio\cite{Ongun2}. It
was reported that single-layer silica has an in-plane stiffness of 22.6 J/m$^2$
which is smaller than that of graphene. Moreover, the Poisson ratio for the
monolayer phase of silica was calculated to be negative like the z-sSiO
monolayer. Having negative Poisson ratio is an important mechanical property
for the usage of a material in biomedical and nanosensor applications.

\section{HALOGENATED SILICENE}\label{Halo}

In order to integrate 2D monolayers into nanotechnological devices, 
fluorination and 
functionalization with other halogen atoms are promising methods similar to 
hydrogenation. 
Experimental and theoretical studies have shown that the band gap of 
fluorinated-graphene can be 
tuned from 0 to 5.0 eV by changing the fluorination 
level\cite{cRobinson,cSamarakoon,cCheng1,cJeon,cGarcia,cWalter1,cWalter2} and 
half-fluorinated graphene was predicted to be magnetic.\cite{cMa} 
Similar to the case of graphene, chemical functionalization of silicene 
with 
fluorine (F) and 
other halogen atoms (such as Cl, Br, and I) have also been extensively studied 
in the literature. 
Gao \textit{et al.} reported that halogenation of silicene opens a band gap 
with 
various gap values 
depending on the 
atomic number of the halogen atoms.\cite{Gao}  The obtained band gaps are 1.19, 
1.47, 1.95, and 1.98 
eV for I, Br, F, and Cl atoms, respectively as shown in Fig. \ref{halog_band}. 
They also reported 
that the formation energy increases with the increase in the atomic number of 
the 
halogen atom. 

As in the hydrogenation case, several structural configurations (see Fig. 
\ref{H-stuc}) were also 
considered for the halogenation of silicene. Ding \textit{et al.} studied the 
structural and 
electronic properties of fluorinated silicene alongside with hydrogenated 
silicene. They 
reported that the band gap of the boat-like (Z-line type in the corresponding 
study) fluorinated 
silicene increases almost linearly with strain, on the other hand, the band 
gap of the 
chair-like structure has a parabolic dependence around the strain value of 
$\epsilon=0.02$.\cite{Ding}

\begin{figure}[htbp]
\includegraphics[width=8cm]{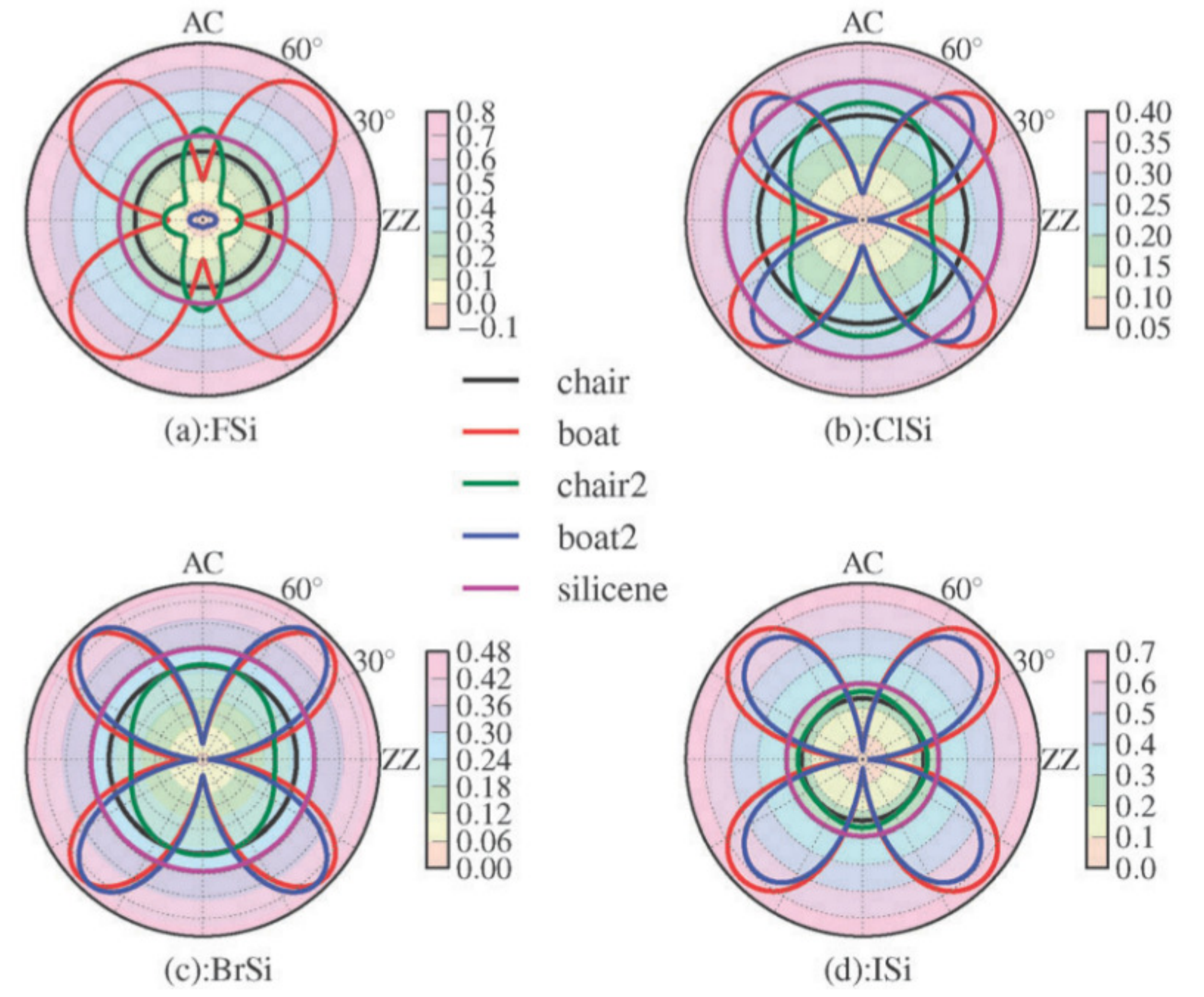}
\caption{\label{halog_poss}
(Color online) Polar diagram for the Poisson ratio, $\nu$, of halogenated
silicene. (a)-(d) Represent the FSi, ClSi, BrSi and ISi systems, respectively. 
The angle $\theta$
identifies the extension direction with respect to the zigzag one. ZZ and AC 
represent
the zigzag and armchair direction, respectively. The numerical values are
represented by different background colors. Isotropic (anisotropic) behavior
is associated with a circular (noncircular) shape of the $\nu_{n}$
plot (taken from Ref. \cite{Zhang2})}
\end{figure}

Zhang \textit{et al.} investigated the geometric and the electronic structure 
as 
well as the 
mechanical properties of halogenated silicene XSi (X = F, Cl, Br and I) in 
various conformers 
(as shown in Fig. \ref{H-stuc}) by using first principles calculations within 
DFT. Their results 
indicated that halogenated silicene shows enhanced stability as compared with 
bare 
silicene and exhibits a 
tunable direct band gap.\cite{Zhang2} They reported that the chair-like 
structure of silicene is 
the most favorable one for all the halogen atoms. They also showed that, 
consistent with the 
previous results\cite{Gao}, the formation energy increases when the atomic 
number of the halogen 
atom increases which indicates that fluorination is the most favorable one 
among all 
halogenation. In addition, as shown in Fig. \ref{halog_poss}, direction 
dependent Poisson 
ratio for different 
conformers of the halogenated silicene were calculated and a negative Poisson 
ratio was predicted for the 
boat-like (boat2 structure in the corresponding study) structure of fluorinated 
silicene\cite{Zhang2}.

Moreover, Wang \textit{et al.} investigated the structural, electronic and 
magnetic
properties of half-fluorinated silicene sheets by using first principles 
simulation within the
framework of DFT. They reported that half-fluorinated (as shown in Fig. 
\ref{H-stuc}) silicene sheets
with zigzag, boat-like or chair-like configurations were confirmed to be 
dynamically stable
based on phonon calculations.\cite{cWang4} Upon the adsorption of fluorine, a 
band 
gap opening is predicted 
in both zigzag and boat-like conformations and they were found to be direct-gap 
semiconductors.
Moreover, half-fluorinated silicene with chair-like configuration shows
antiferromagnetic ordering which is mainly induced by the unfluorinated Si 
atoms. 

Wei \textit{et al.} investigated the optical properties of fluorinated silicene 
by using 
the many-body effects by using Green’s function perturbation
theory.\cite{Wei} As in hydrogenation, fluorination of silicene also opens a 
band 
gap which is 
consistent with the previous studies. They also reported that strong 
excitonic effects 
dominate the absorption properties of hydrogenated, fluorinated silicene, and 
silicene 
nanoribbon with high exciton binding energies.

\section{Functionalization via Organic Molecule Adsorption}\label{Organic}
The adsorption of different chemical functional groups on silicene
have potential applications for silicene-based nanoelectronic devices.
Different from the highly stable planar structure of graphene, the buckled 
honeycomb
structure of silicene leads to high chemical reactivity for functional
groups. Thus, adsorption of functional groups could be a prominent method
for tuning the electronic structure of silicene.

\begin{figure}[htbp]
\includegraphics[width=8.0cm]{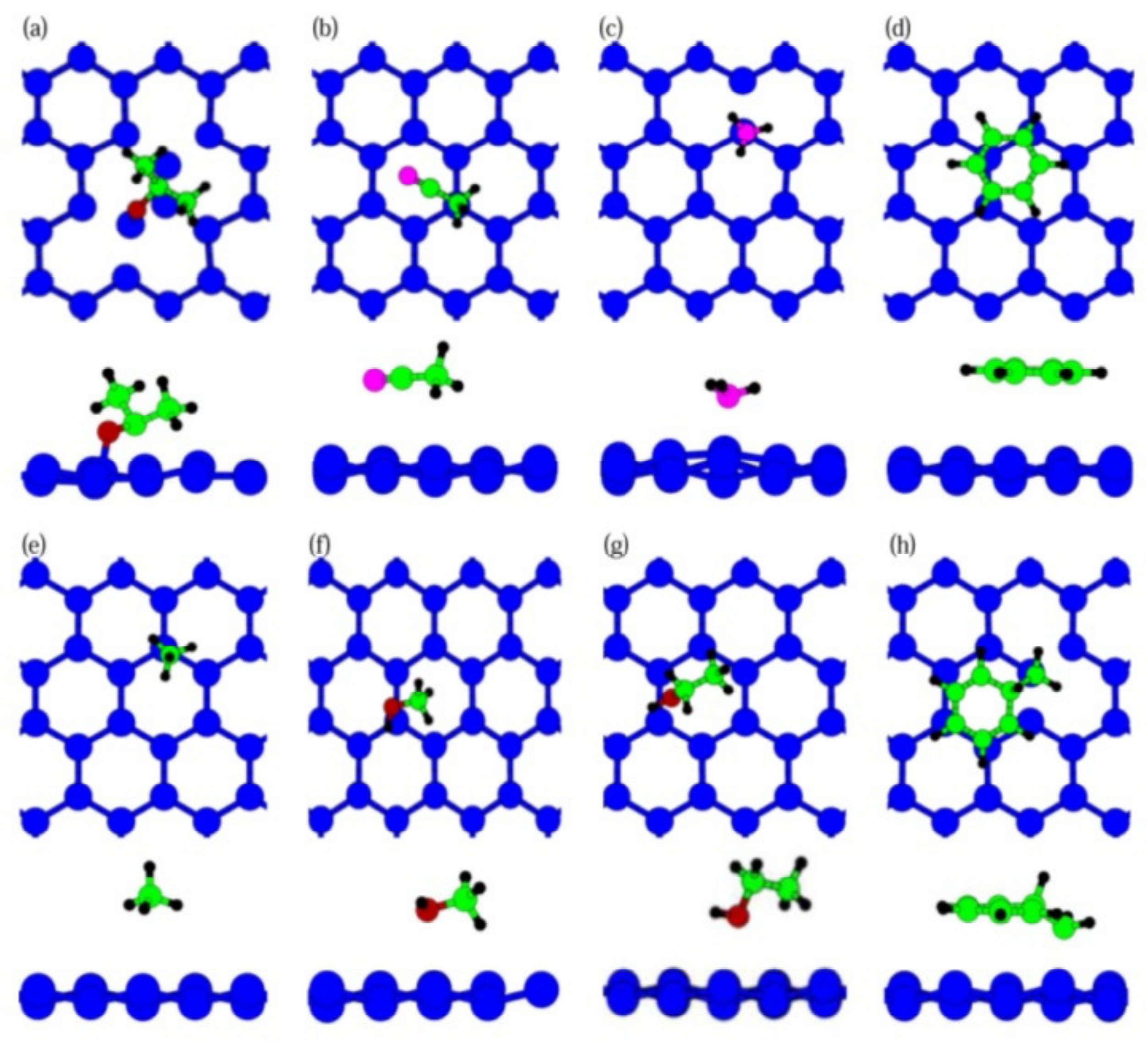}
\caption{\label{various-atoms}
(Color online) Top and side views of relaxed geometric structures of (a) 
acetone, (b) acetonitrile, 
(c) ammonia, (d) benzene, (e) methane, (f) methanol, (g) ethanol, and (h) 
toluene molecules on 
silicene. Blue, green, black, violet, and red spheres are Si, C, H, N, and O 
atoms,
respectively (taken from Ref. \cite{Kaloni})}
\end{figure}

Hue \textit{et al.} investigated the adsorption of NH$_{3}$, NO and NO$_{2}$ on 
silicene and found that the electronic 
properties of
silicene are strongly depend on the type of adsorbate. Their findings revealed 
a 
significant
potential of silicene for highly sensitive molecule sensors. In addition, Wen 
\textit{et al.} found high reactivity of silicene towards NO$_{2}$, 
O$_{2}$ and SO$_{2}$ 
molecules.\cite{Feng} Binding energies of these molecules on silicene are 
larger
than 1 eV. In contrast, the binding energies of NO and NH$_{3}$ are 0.35 and 
0.60 eV, 
respectively. While
the band gap of silicene is enhanced upon adsorption of NO, O$_{2}$, NH$_{3}$, 
and
SO$_{2}$, it becomes half-metallic when NO$_{2}$ is adsorbed. The structural 
and 
electronic
properties of diverse molecules adsorbed on silicene were investigated by van 
der 
Waals included
DFT.\cite{Kaloni} Considered molecules are shown in Fig. \ref{various-atoms} 
and their calculated
adsorption energies vary from -0.11 to -0.95 eV indicating no adsorption. 
Moreover, electronic structure in hydrogenated
silicene as well as fluorinated silicene 
calculations showed that
the calculated band gaps range from 0.01 to 0.35 eV for acetonitrile to 
acetone, respectively.

Recently, Prasongkit \textit{et al.} investigated the change of the electronic 
and transport 
properties when NO$_{2}$, NO, NH$_{3}$, and CO molecules are adsorbed onto 
pristine and B/N-doped 
silicene.\cite{Prasongkit} Their 
results showed that NO and
NO$_{2}$ can be sensitively detected by pristine silicene. On the other hand, 
due to the weak interaction
of CO and NH$_{3}$ molecules with pristine silicene, the possibility of 
detection of those gases is 
relatively low. Increased sensitivity toward NH$_{3}$ and CO obtained when 
pristine silicene is doped 
either by B or N atoms. Quantum conductance properties of CO molecule adsorbed 
silicene nanoribbons were 
investigated by Osborn \textit{et al.}\cite{Osborn1} They showed that the 
quantum 
conduction is modified in a detectable way by weak chemisorption of a single CO 
molecule on a silicene 
nanoribbon. The adsorption of N$_{2}$ and 
CO$_{2}$ molecules do not affect the conductance. However, O$_{2}$ and H$_{2}$O 
molecules can be strongly 
chemisorbed 
and can diminish the 
CO detection capability of silicene. Moreover, they found that CO, O$_{2}$ and 
H$_{2}$O are easily 
detectable molecules among CO, CO$_{2}$, O$_{2}$, N$_{2}$, and H$_{2}$O. Gurel 
\textit{et al.} investigated the 
interaction of H$_{2}$, O$_{2}$, CO, H$_{2}$O, and OH molecules with graphene 
and silicene\cite{Gurel} and found that H$_{2}$, O$_{2}$, and CO remain intact 
on both 
graphene and silicene. 
When these molecules adsorb at the vicinity of vacancy centers they can 
dissociate. The dissociations 
of other atoms are hindered by high energy barriers. Stephan \textit{et al.} 
studied 
adsorption of benzene molecule on 
($3\times3$) silicene which was placed on the ($4\times4$) Ag (111) 
surface.\cite{Stephan} Their 
study revealed that
benzene molecule can be chemisorbed on a silicene layer deposited on Ag(111) 
through a cycloaddition reaction.
They also showed that, Si (100) and Si (111) surfaces are more reactive than 
the other surfaces of the structure.
In addition, Stephan \textit{et al.} investigated the adsorption 
characteristics 
of H$_{2}$Pc molecule on silicene above Ag (111).\cite{Stephan1} They 
showed 
that, due to an
electostatic or polarization repulsion between H$_{2}$Pc molecule and Si 
surface,
H$_{2}$Pc molecule adopts a butterfly configuration on this surface. However, 
this molecule shows a planar
configuration on the SiC and SiB surfaces. This study revealed the possibility 
of chemisorption of such large
molecules on the Si/Ag system.

\section{Functionalization of silicene via adatom decoration}\label{doping}
Due to the buckled honeycomb structure of silicene, it is chemically a very 
active material.
In order to maintain and tune its electronic properties as required, diverse 
growth mechanisms 
and various substrates were used. During growth processes the presence of 
foreign atoms and cluster formation is inevitable. The quality of fabricated 
silicene-based 
devices is strongly 
affected by the adsorbed foreign atoms. Therefore, the investigation of the 
decoration mechanisms of these atoms 
on silicene is quite essential.

\begin{figure}[htbp]
\includegraphics[width=8.0cm]{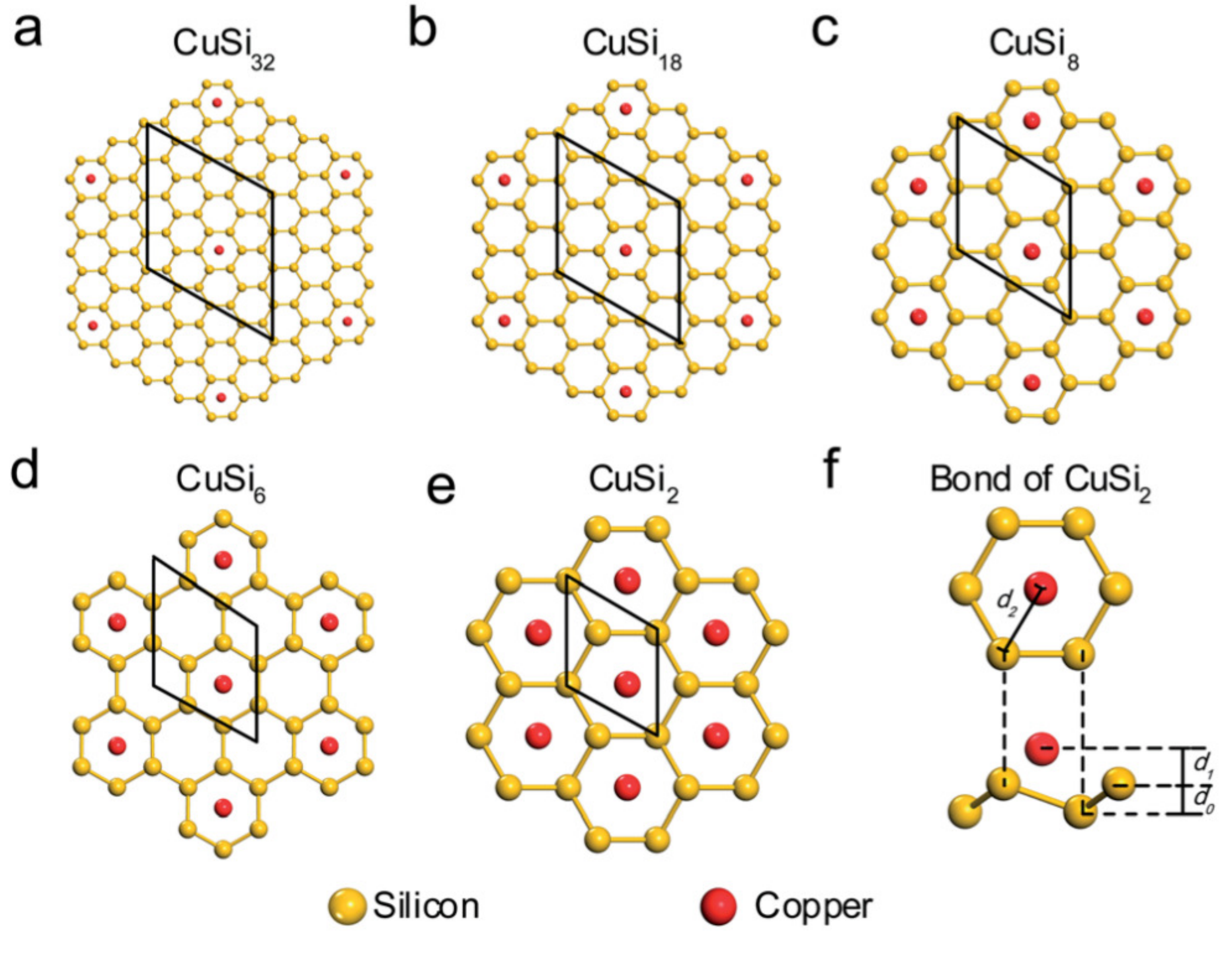}
\caption{\label{fig-9}
(Color online) Optimized geometric structures of Cu-covered silicene with a 
coverage 
of N = 3.1\% (a), 5.6\% (b), 12.5\% (c), 16.7\% (d), and 50.0\% (e). (f) Top 
and side views of a Cu-covered silicene supercell. The 
parallelograms show the unit cell for each structure and the parameters 
d$_{0}$ and d$_{1}$ demonstrate the buckling of siilicene and the height of Cu 
atom to upper Si atom, respectively. The distance d$_{2}$ represents the 
distance from the adsorbed atom to the upper Si atom in silicene (taken from 
Ref. \cite{Ni})}
\end{figure}

\begin{figure}[htbp]
\includegraphics[width=8.0cm]{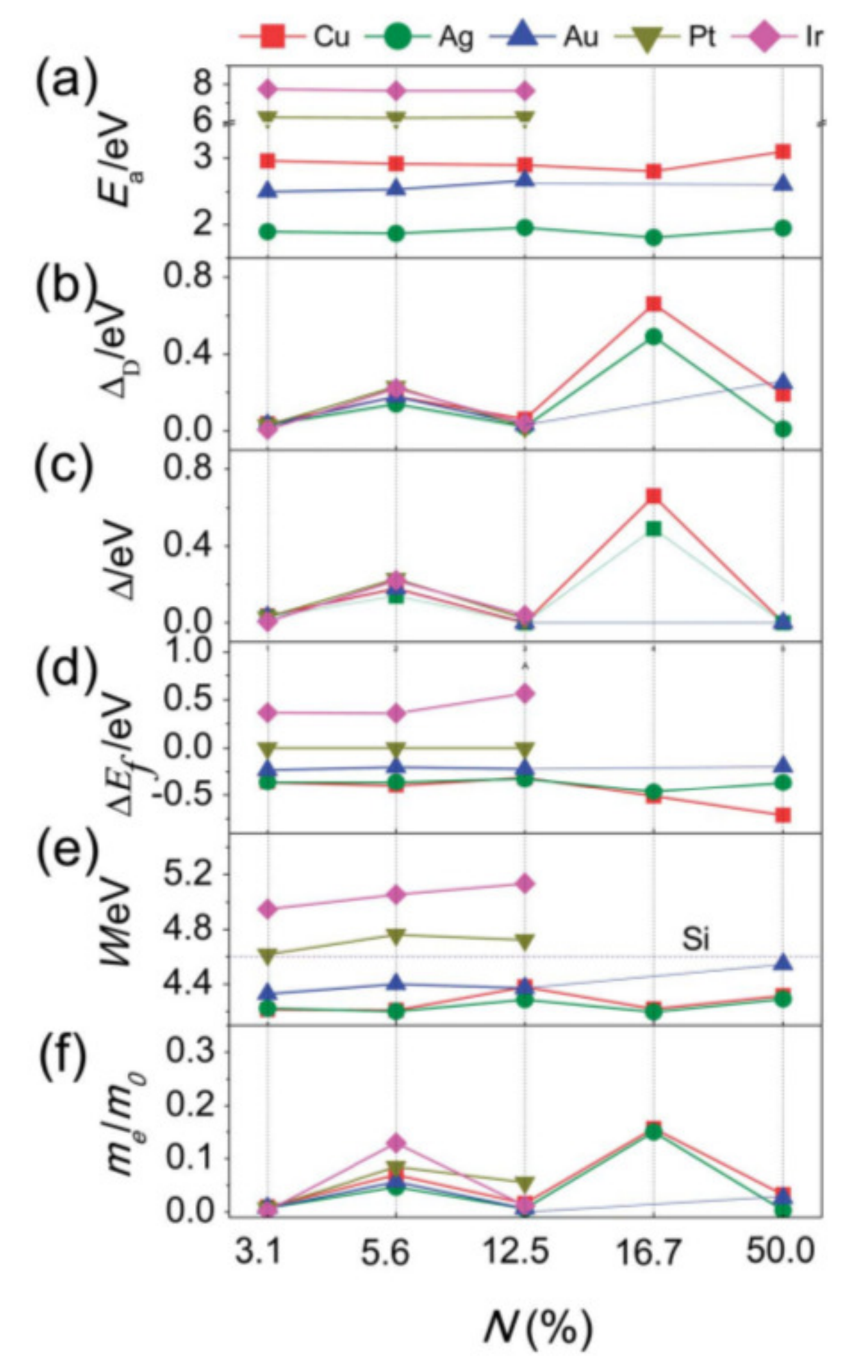}
\caption{\label{fig-10}
(Color online)  Calculated adsorption energy (per atom) (a), band gap at the 
Dirac point (b), global band gap (c), Fermi level shift of metal covered-
silicene (d), work function (the horizontal dashed line shows the work 
function of pure silicene) (e),  and effective mass of holes of the
metal covered silicene (f) as a function of coverag (taken from Ref. \cite{Ni})}
\end{figure}

Ni \textit{et al.} investigated the geometric and electronic properties of 
silicene with five different 
transition metal atoms (Cu, Ag, Au, Pt, and Ir) adsorbed at different coverages 
by using first principle 
methods.\cite{Ni} Optimized geometric structures of Cu-covered silicene 
with different covarages are shown 
in Fig. \ref{fig-9}. Similiar to Cu, favorable geometric configuration of Ag , 
Au, Pt, and Ir atoms is the center of silicene hegzagons. A sizable band gap 
can be opened without 
degrading the electronic properties at 
the Dirac point of silicene when these atoms are adsorbed. Adsorption 
characteristics of the metal atoms are given in Fig. \ref{fig-10}. As shown in 
the 
figure, a band gap opening occurs in all the considered coverages and the 
value of the gap increases from 0.03 to 0.66 eV with 
increasing coverage range. Using the method of the 
adsorption of different transition 
metal atoms on different regions of silicene, they designed a silicene p-i-n 
tunnelling field effect transistor.

\begin{figure}[htbp]
\includegraphics[width=8.0cm]{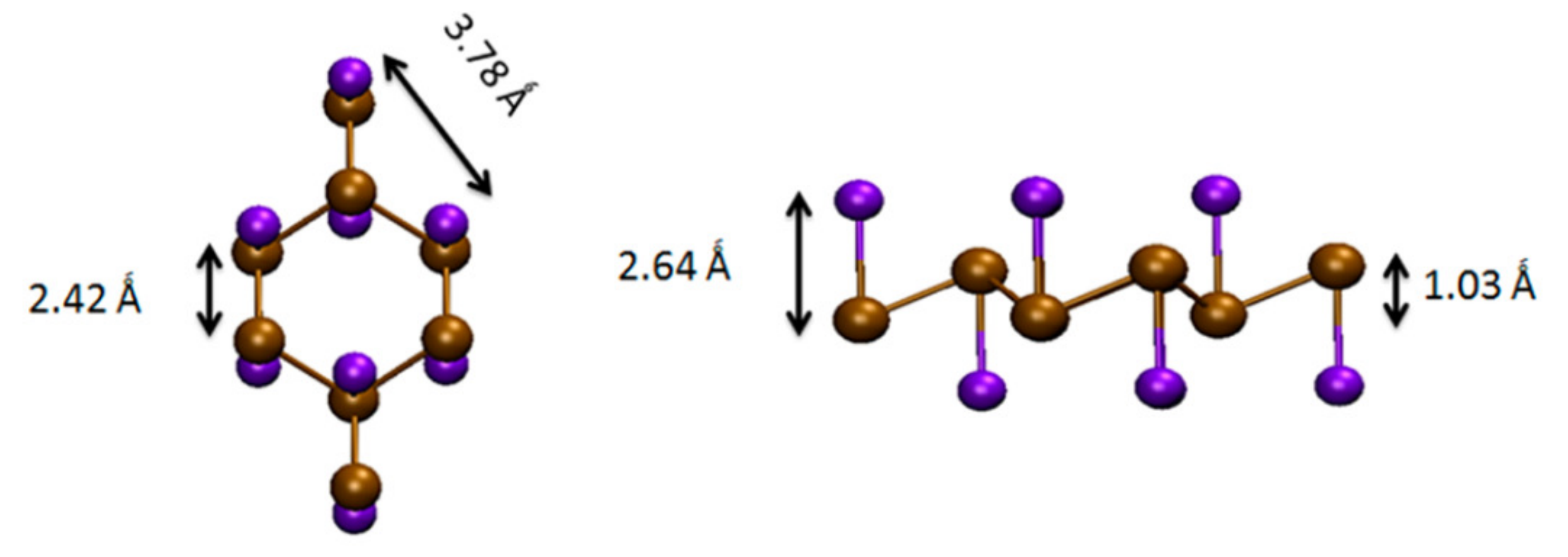}
\caption{\label{fig-11}
(Color online) Optimized geometric structure of fully lithiated silicene
(silicel) (taken from Ref. \cite{Osborn2})}
\end{figure}

The adsorption characteristics and the stability of Li atoms on silicene was
investigated by first principles calculations.\cite{Osborn2, Tritsaris} It was  
reported that, Li adsorbed silicene compounds are energetically favorable and  
fully lithiated silicene (silicel) is the most stable form among them.
The stability of the silicene sheet in the presence of completely adsorbed 
lithium atoms on
the atom-down sites of both sides (Fig. \ref{fig-11}) was confirmed by
molecular dynamic simulations conducted at elevated temperatures. Lithiation
can be used to tune the band gap of silicene and complete Li adsorption results 
in a band gap of 0.368 eV 

Quhe \textit{et al.} examined the gap opening in silicene 
in the presence of single-side adsorption of alkali atoms such as Li, Na, K, 
Rb, and 
Cs.\cite{Quhe} 
They showed that the band gap of silicene can be tuned by alteration of 
adsorption coverage resulting in a band gap up to 0.5 eV. Moreover, quantum 
transport simulation of a 
bottom-gated FET 
based on a Na-covered silicene was also conducted and a transport 
gap with an on/off current ratio up to  10$^{8}$ was predicted. 
The electronic structure, mechanical stability, and hydrogen storage capacity 
of strain induced Mg functionalized 
silicene (SiMg) and silicane (SiHMg) monolayers have been investigated by 
Hussain \textit{et 
al.}\cite{Hussain} Their results revealed that high doping concentration of Mg 
atom can be 
obtained on both monolayers by biaxial symmetric strain up to 10\%. The 
adsorption energy 
of H$_{2}$ molecules on silicene was found to be ideal for the application in 
hydrogen storage devices.
Li, Na, K, Be Mg and Ca adsorbed silicene sheets were studied to investigate 
their 
hydrogen-storage capacity.\cite{Hussain1} It is found that Li and Na atoms have 
strong 
metal-to-substrate binding and they are suitable for high-capacity storage 
of hydrogen. Using 
DFT calculations, the effects of an external electric field on the 
adsorption-desorption of H$_{2}$ on a Ca-decorated silicene system was 
studied.\cite{Song1} Doubled 
binding energy enhancement is observed for H$_{2}$ when 0.004 au external 
electric field was applied on the Ca-silicene 
system. On the other hand, when -0.004 au external electric field was applied 
to 
the system, the binding of 9H$_{2}$ on Ca- monolayer or bilayer silicene system 
is getting  
weaker.

\section{Summary}\label{conc}
Functionalization of 2D materials is an efficient way to tailor their 
electronic, optical and mechanical properties. Silicene, 
with its highly-reactive surface structure, is a good candidate for various functionalization techniques 
and have been widely studied by the researchers. Recent studies have demonstrated 
that hydrogenation and halogenation of silicene can tune the electronic-band 
structure from semi-metal to semiconductor. However, by single-side adsorption 
of alkali atoms, Li, Na, K, Rb, and Cs, only a relatively 
small 
band gap opens. Opening a gap in silicene is rather important for its potential 
usage 
of the material in optoelectronic applications. 
In addition to its electronic properties, chemical functionalization of 
silicene 
can also change the Poisson ratio from positive to negative values which is 
important 
for 
applications in biomedicine and nanosensors. Like other 2D monolayer materials, 
the electronic, optical and mechanical properties of silicene could be tuned by 
chemical functionalization to integrate it into 
nanotechnological device applications.

\begin{acknowledgments}

This work was supported by the Flemish Science Foundation (FWO-Vl) and the
Methusalem foundation of the Flemish government. Computational resources were
provided by TUBITAK ULAKBIM, High Performance and Grid Computing Center
(TR-Grid e-Infrastructure). H.S. is supported by a FWO Pegasus Long Marie Curie
Fellowship.

\end{acknowledgments}


\begin{thebibliography}{99}


-----------------------------------------

\bibitem{Brodie} B. C. Brodie,
Phil. Trans. R. Soc. Lond. \textbf{149}, 249 (1859).

\bibitem{Peierls1} R. E. Peierls, Ann. Inst. Henri Poincare
\textbf{5}, 177 (1935).

\bibitem{Peierls2} R. E. Peierls, Helv. Phys. Acta,
\textbf{7}, 81 (1934).

\bibitem{Novo1} K. S. Novoselov, A. K. Geim, S. V. Morozov, D. Jiang, Y. Zhang,
S. V.
Dubonos, I. V. Grigorieva, and A. A. Firsov,
Science \textbf{306}, 666 (2004).

\bibitem{Novo2} K. S. Novoselov, D. Jiang, F. Schedin, T. Booth, V. V. 
Khotkevich, S. Morozov, and A. K. Geim, Proc. Natl. Acad. Science
U.S.A. \textbf{102}, 10451 (2005).

\bibitem{Geim1} K. S. Novoselov, A. K. Geim, S. V. Morozov, D. Jiang, M. I.
Katsnelson,
I. V. Grigorieva, S. V. Dubonos, and A. A. Firsov,
Nature \textbf{438}, 197 (2005).

\bibitem{hasan1} H. Sahin, S. Changirov, M. Topsakal, E. Bekaroglu, E. Akturk,
R. T. Senger, and S. Ciraci, Phys. Rev. B \textbf{80}, 155453 (2009).

\bibitem{Golberg} D. Golberg, Y. Bando, Y. Huang, T. Terao,
M. Mitome,
C. Tang, and C. Zhi,
ACS Nano \textbf{4}, 2979 (2010).

\bibitem{Zeng} H. Zeng, H. Zhi, C. Zhang, Z. Wei, X. Wang, X. Guo, W. Bando, Y.
Golberg, D. Nano Lett. \textbf{10}, 5049
(2010).

\bibitem{Song} L. Song, L. Ci, L. Lu, H. Sorokin, P. B. Jin, C. Ni, J.
Kvashnin,
A. G. Kvashnin, D. G. Lou, J. Yakobson, B. I. Ajayan, P. M. Nano Lett.
\textbf{10}, 3209
(2010).

\bibitem{Bacaksiz} C. Bacaksiz, H. Sahin, H. D. Ozaydin, S. Horzum, R. T.
Senger, and F. M. Peeters, Phys. Rev. B \textbf{91}, 085430
(2015).

\bibitem{Zhuang} H. L. Zhuang and R. G. Hennig, Appl. Phys. Lett. \textbf{101},
153109
(2012).

\bibitem{QWang} Q. Wang, Q. Sun, P. Jena, and Y. Kawazoe, ACS Nano \textbf{3},
621
(2009).

\bibitem{KKim} K. K. Kim, A. Hsu, X. Jia, S. M. Kim, Y. Shi, M. Hofmann, D.
Nezich, J. F. Rodriguez-Nieva,
M. Dresselhaus, T. Palacios, and J. Kong, Nano Lett. \textbf{12}, 161
(2012).

\bibitem{MFarahani} M. Farahani, T. S. Ahmadi, and A. Seif, J. Mol.
Struct. \textbf{913}, 126
(2009).

\bibitem{Wang} Q. H. Wang, K. K. Zadeh, A. Kis, J. N. Coleman, and M. S.
Strano,
Nat. Nanotechnol. \textbf{699}, 699 (2012).

\bibitem{Wilson} J. A. Wilson and A. D. Yoffe, Adv. Phys.  \textbf{18}, 193
(1969).

\bibitem{Horzum} S. Horzum, D. Cakir, J. Suh, S. Tongay, Y. S. Huang, C. H. Ho, 
J. Wu, H. Sahin, and F. M. Peeters, Phys. Rev. B \textbf{89}, 155433 (2014).

\bibitem{Bacaksiz2} C. Bacaksiz, S. Cahangirov, A. Rubio, R. T. Senger, F. M. 
Peeters, and H. Sahin, Phys. Rev. B \textbf{93}, 125403
(2016).

\bibitem{Cahangirov} S. Changirov, M. Topsakal, E. Akturk, H. Sahin, and S.
Ciraci, Phys. Rev. B \textbf{102}, 236804 (2009).

\bibitem{Vogt} P. Vogt, P. D. Padova, C. Quaresima, J. Avila, E. Frantzeskakis,
M. C. Asensio, A. Resta, B. Ealet, and G. L. Lay, Phys. Rev. Lett. \textbf{108},
155501 (2012).

\bibitem{Lin} C. L. Lin, R. Arafune, K. Kawahara, N. Tsukahara, E. Minamitami,
Y. Kim, N. Takagi, and M. Kawai,
Appl.
Phys. Express \textbf{5}, 045802
(2012)

\bibitem{Fleurence} A. Fleurence, R. Friedlein, T. Osaki, H. Kawai, Y. Wang,
and Y. Y. Takamura, Phys. Rev. Lett.
\textbf{108}, 245501 (2012).

\bibitem{Davila} M. E. Davila, L. Xian, S. Cahangirov, A. Rubio, and G. L. Lay,
 New J. Phys. \textbf{16} 095002 (2014).

\bibitem{Zhu} F. F. Zhu, W. J. Chen, C. L. Gao, D. D. Guan, C. H.
Liu, D. Qian, S. C. Zhang, and J. F. Jia,
 Nat. Mater. \textbf{14} 1020 (2015).


\bibitem{Boukhvalov} D. W. Boukhvalov, M. I. Katsnelson, and A. I. 
Lichtenstein, 
Phys. Rev. B
\textbf{77}, 035427 (2008).

\bibitem{Haberer} D. Haberer \textit{et al.}, Nano Lett. \textbf{10}, 3360 
(2010).

\bibitem{Sofo} J. O. Sofo, A. S. Chaudhari, and G. D. Barber, Phys. Rev. B 
\textbf{75}, 153401 (2007).

\bibitem{cRobinson} J. T. Robinson, J. S. Burgess, C. E. Junkermeier,
S. C. Badescu, T. L. Reinecke, F. K. Perkins,
M. K. Zalalutdniov, J. W. Baldwin, J. C. Culbertson,
P. E. Sheehan, and E. S. Snow, Nano Lett.  \textbf{10}, 3001 (2010).


\bibitem{cSamarakoon} D. K. Samarakoon, Z. Chen, C. Nicolas, and X. Q. Wang,
Small \textbf{7}, 965 (2011).

\bibitem{cCheng1} S. H. Cheng, K. Zou, F. Okino, H. R. Gutierrez, A. Gupta,
N. Shen, P. C. Eklund, J. O. Sofo, and J. Zhu, Phys. Rev. B \textbf{81}, 205435
(2010).

 \bibitem{cJeon}  K. J. Jeon, Z. Lee, E. Pollak, L. Moreschini, A. Bostwick,
C. M. Park, R. Mendelsberg, V. Radmilovic, R. Kostecki,
T. J. Richardson, and E. Rotenberg, ACS Nano  \textbf{5}, 1042 (2011)


\bibitem{cGarcia} J. C. Garcia, D. D. B. Lima, L. V. C. Assali, and J. F. Justo,
J.
Phys. Chem. C \textbf{115}, 13242 (2011).

\bibitem{cWalter1} A. L. Walter, H. Sahin, K. J. Jeon, A. Bostwick, S. Horzum, 
R. Koch, 
F. Speck, M. Ostler, P. Nagel, M. Merz, S. Schupler, L. Moreschini, 
Y. J. Chang, T. Seyller, F. M. Peeters, K. Horn, and E. Rotenberg, ACS Nano 
\textbf{8}, 
7801 (2014).


\bibitem{cWalter2} A. L. Walter, H. Sahin, J. Kang, K. J. Jeon, A. Bostwick, S. 
Horzum, 
L. Moreschini, Y. J. Chang, F. M. Peeters, K. Horn, and E. Rotenberg,
Phys. Rev. B \textbf{93}, 075439 (2016).

 \bibitem{Kara} A. Kara, H. Enriquez, A. P. Seitsonen, L. C. L. Y. Voon, S.
Vizzini, B.
Aufray, and H. Oughaddou,
 Surf. Sci. Rep. \textbf{67}, 1 (2012).

\bibitem{Xu1} M. Xu, T. Liang, M. Shi, and H. Chen,
 Chem. Rev. \textbf{113}, 3766 (2013).
 
 \bibitem{hasan2} H. Sahin, J. Sivek, S. Li, B. Partoens, and F. M. Peeters
Phys. Rev. B \textbf{88}, 045434 (2013).
 

\bibitem{Lalmi} B. Lalmi, H. Oughaddou, H. Enriquez, A. Kara, S.
Vizzini, B. Ealet, and B. Aufray, Appl. Phys. Lett. \textbf{97} 223109 (2010).

\bibitem{Meng} L. Meng, Y. Wang, L. Zhang, S. Du, R. Wu, L. Li,
Y. Zhang, G. Li, H. Zhou, W. A. Hofer, and  H. J. Gao, Nano Lett.
\textbf{13} 685 (2013).

\bibitem{Lin2} X. Lin and J. Ni,
 Phys. Rev. B \textbf{86}, 075440 (2012).

\bibitem{Quhe} R. Quhe, R. Fei, Q. Liu, J. Zheng, H. Li, C. Xu, Z. Ni, Y. Wang,
D. Yu, Z. Gao, and J.
Lu, Sci. Rep. \textbf{2}, 853 (2012).

\bibitem{Ni} Z. Ni, H. Zhong, X. Jiang, R. Quhe, G. Luo, Y. Wang, M. Ye, J.
Yang, J. Shi, and J. Lu, Nanoscale
\textbf{6}, 7609 (2014).

\bibitem{Cheng} Y. C. Cheng, Z. Y. Zhu, and U. Schwingenschlogl,
 Europhys. Lett. \textbf{95}, 17005 (2011).

\bibitem{Sivek} J. Sivek, H. Sahin, B. Partoens, and F. M. Peeters, Phys. Rev.
B \textbf{87}, 085444 (2013).

\bibitem{Zheng} R. Zheng, X. Lin, and J. Ni, Appl. Phys. Lett.
\textbf{105}, 092410 (2014).

\bibitem{Okamoto1} H. Okamoto, Y. Sugiyama, and H. Nakano, Chem. Eur. J.
\textbf{17}, 9864 (2011).

\bibitem{Nakano} H. Nakano, M. Nakano, K. Nakanishi, D. Tanaka, Y. Sugiyama, T.
Ikuno,
H. Okamoto, and T. Ohta, J. Amer. Chem. Soc. \textbf{134}, 5452 (2012).

\bibitem{Okamoto2} H. Okamoto, Y. Kumai, Y. Sugiyama, T. Mitsuoka, K.
Nakanishi, T. Ohta,
 H. Nozaki, S. Yamaguchi, S. Shirai, and H. Nakano, J. Amer. Chem. Soc.
\textbf{132}, 2710 (2010).

\bibitem{Sugiyama} Y. Sugiyama, H. Okamoto, T. Mitsuoka, T. Morikawa, K.
Nakanishi, T. Ohta, and
H. Nakano, J. Amer. Chem. Soc. \textbf{132}, 5946 (2010).


\bibitem{Pereda} P. R. Pereda and N. Takeuchi, J. of Chem. Phys. \textbf{138},
194702 (2013).

\bibitem{Spencer} M. J. S. Spencer, M. R. Bassett, T. Morishita,  I. K. Snook,
and H. Nakano, New J. Phys. \textbf{15}, 125018 (2013).

\bibitem{Du} Y. Du, J. C. Zhuang, H. S. Liu, X. Xu, S. Eilers, K. H. Wu, C.
Peng, J. J. Zhao, X. D. Pi,
K. See, G. Peleckis, X. Wang, and X. Dou, ACS Nano \textbf{8}, 10019 (2014).

 \bibitem{Liu12} G. Liu, M. S. Wu, C. Y. Ouyang, and B. Xu, Euro Phys. Lett. 
\textbf{99}, 17010 (2012).

 \bibitem{Qin12} R. Qin, C. H. Wang, W. Zhu, and Y. Zhang, AIP Adv. \textbf{2},
022159 (2012).

\bibitem{Zhao12} H. Zhao, Phys. Lett. A \textbf{376}, 3546 (2012).

\bibitem{Hu13} M. Hu, X. Zhang, and D. Poulikakos, Phys. Rev. B \textbf{87},
195417 (2013).

\bibitem{Kal-13} T. P. Kaloni, Y. C. Cheng, and U. Schwingenschlögl, J. App.
Phys. \textbf{113}, 104305 (2013).

 \bibitem{Dur14} A.P. Durajski, D. Szczesniak, and R. Szczesniak, Solid State
Comm. \textbf{200}, 17 (2014).

 \bibitem{Moh14} B. Mohan, A. Kumar, and P.K. Ahluwalia, Physica E \textbf{61}, 
40
(2014).

 \bibitem{Hus14} T. Hussain, S. Chakraborty, A. D. Sarkar, B. Johansson, and
R. Ahuja, J. App.
Phys. \textbf{105}, 123903
(2014).

\bibitem{Wang14} B. Wang, J. Wu, X. Gu, H. Yin, Y. Wei, R. Yang, and M.
Dresselhaus, Appl. Phys. Lett. \textbf{104}, 081902
(2014).

\bibitem{Zhu14} J. Zhu and U. Schwingenschlogl, ACS Appl. Mater. Interfaces
\textbf{6}, 11675
(2014).


\bibitem{Yang14} C. Yang, Z. Yu, P. Lu, Y. Liu, H. Ye, and T. Gao, Comp.
Mater. Sci.
\textbf{95}, 420
(2014).

\bibitem{Cao15} G. Cao and Y. Zhang, J. Cao, Phys. Lett. A
\textbf{379}, 1475
(2015).

\bibitem{Ding} Y. Ding and Y. Wang, Appl. Phys. Lett. \textbf{100}, 083102
(2012).

\bibitem{Voon} L. C. L. Y. Voon, E. Sandberg, R. S. Aga, and A. A. Farajian,
Appl. Phys.
Lett. \textbf{97}, 163114
(2010).


\bibitem{Houssa} M. Houssa, E. Scalise, K. Sankaran, G. Pourtois, V. V.
Afanasev, and A. Stesmans, Appl. Phys. Lett.
\textbf{98}, 223107
(2011).

\bibitem{Zhang} P. Zhang, X. D. Li, C. H. Hu, S. Q. Wu, and Z. Z. Zhu, Phys.
Lett. A
\textbf{376}, 1230
(2012).

\bibitem{Wei} W. Wei and T. Jacob, Phys. Rev. B \textbf{88}, 045203
(2013).

\bibitem{Gao} N. Gao, W. T. Zheng, and Q. Jiang, Phys. Chem. Chem. Phys.
\textbf{14}, 257
(2012).

\bibitem{Zhang2} W. B. Zhang, Z. B. Song, and L. M. Dou, J. Mater. Chem. C
\textbf{3}, 3087
(2015).

\bibitem{Wang2} X. Wang, H. Liu and S. T. Tu, RSC Adv. \textbf{36}, 6238
(2015).



 \bibitem{Padova1} P. De Padova, C. Ottaviani, C. Quaresima, B. Olivieri, P. 
Imperatori, E.
Salomon,
 T. Angot, L. Quagliano, C. Romano, A. Vona,  M. M. Miranda, A. Generosi,
 B. Paci, and G. L. Lay, 2D Mater. \textbf{1}, 021003
(2014).

\bibitem{Padova2} P. De Padova, C. Quaresima, B. Olivieri, P. Perfetti, and G.
L. Lay, J. Phys.
D: Appl. Phys.
\textbf{44}, 312001
(2011).

\bibitem{Molle} A. Molle, C. Grazianetti, D. Chiappe, E. Cinquanta, E. Cianci,
G. Tallarida, and
 M. Fanciulli, Adv. Func. Mater.
\textbf{24}, 5088
(2013).

\bibitem{Friedlein} R. Friedlein, H. V. Bui, F. B. Wiggers, Y. Y. Takamura, A.
Y. Kovalgin,
and M. P. de Jong, J. Chem. Phys. \textbf{140}, 204705
(2014).

\bibitem{Xu} X. Xu, J. Zhuang, Y. Du, H. Feng, N. Zhang, C. Liu, T. Lei, J.
Wang, M. Spencer,
 T. Morishita, X. Wang, and S. X. Dou, Sci. Rep. \textbf{4}, 7543
(2014).


\bibitem{Liu} G. Liu, X. L. Lei, M. S. Wu, B. Xu, and C. Y. Ouyang, J. Phys.
Conden.
Matter. \textbf{26}, 355007
(2014).

\bibitem{Drissi1} L. B. Drissi, E. H. Saidi, M. Bousmina, and O. F. Fehri, J.
Phys. Conden.
Matter. \textbf{24}, 485502
(2012).

\bibitem{Drissi2} L. B. Drissi and F. Z. Ramadan, Physica E \textbf{68}, 38
(2015).

\bibitem{Zhang3} P. Zhang, B. B. Xiao, X. L. Hou, Y. F. Zhu, and Q. Jiang, Sci.
Rep. \textbf{4}, 3821
(2014).

\bibitem{Osborn} T. H. Osborn, A. A. Farajian, O. V. Pupysheva, R. S. Aga, and
L.C. L. Y. Voon , Chem. Phys. Lett.
\textbf{511}, 101
(2011).

\bibitem{Zhang4} C. W. Zhang and S. S. Yan, J. Phys. Chem. C \textbf{116}, 4163
(2012).

\bibitem{Sahin3} H. Sahin and F. M. Peeters, Phys. Rev. B \textbf{87}, 085423
(2013).


\bibitem{Chinnathambi} K. Chinnathambi, A. Chakrabarti, A. Banerjee, and S.K.
Deb, arXiv:1205.5099v1.

\bibitem{Rong} W. Rong, M. S. Xu, and X. D. Pi, Chinese Phys. B \textbf{24}, 086807
(2015).

\bibitem{Huang} B. Huang, H. X. Deng, H. Lee, M. Yoon, B. G. Sumpter, F. Liu,
S. C. Smith, and S. H. Wei, Phys. Rev. X \textbf{4}, 021029
(2014).

\bibitem{Liu2} Y. Liu, H. Shu, P. Liang, D. Cao, X. Chen, and W. Lu, J. Appl. 
Phys. \textbf{114}, 094308
(2013).

\bibitem{Jamdagni} P. Jamdagni, A. Kumar, M. Sharma, A. Thakur, and P. K.
Ahluwalia, AIP Conf. Proc. \textbf{1661}, 080007
(2015).

\bibitem{Peng} Q. Peng and S. De, Nanoscale \textbf{6}, 1207
(2014).

\bibitem{Yang} C. H. Yang, Z. Y. Yu, P. F. Lu, Y. M. Liu, S. Manzoor, M. Li,
and S. Zhou, Proc. SPIE \textbf{8975}, 89750K-1
(2014).

\bibitem{Wang3} Y. Wang and Y. Ding, Phys. Stat. Solidi RRL \textbf{7}, 410
(2013).

\bibitem{Ongun1} V. O. Ozcelik and S. Ciraci, J. Phys. Chem. C \textbf{117},
26305
(2013).

\bibitem{Ongun2} V. O. Ozcelik, S. Cahangirov, and S. Ciraci, Phys. Rev. Lett.
\textbf{112}, 246803
(2014).

\bibitem{Gurel} H. H. Gurel, V. O. Ozcelik, and S. Ciraci, J. Phys. Chem. C
\textbf{118}, 27574
(2014).


\bibitem{cMa} Y. D. Ma, Y. Dai, M. Guo, C. W. Niu, L. Yu, and B. B. Huang,
Nanoscale \textbf{3}, 2301  (2011).

\bibitem{cWang4} X. Wang, H. Liu, and S. T. Tu, RSC Adv.  \textbf{5}, 6238
(2015).

\bibitem{Hu} W. Hu, N. Xia, X. Wu, Z. Li, and J. Yang, Phys. Chem. Chem. Phys.
\textbf{16}, 6957 (2014).
\bibitem{Feng} J. Wen, Y. J. Liu, H. X. Wang, J. X. Zhao, Q. H. Cai, and X. Z.
Wang,
Comp. Mater. Sci. \textbf{87}, 218 (2014).
\bibitem{Kaloni} T. P. Kaloni, G. Schreckenbach, and M. S. Freund, J. Phys.
Chem. C
\textbf{118}, 23361 (2014).
\bibitem{Prasongkit} J. Prasongkit, R. G. Amorim, S. Chakraborty, R. Ahuja, R.
H. Scheicher,
and V. Amornkitbamrung, J. Phys. Chem. C
\textbf{119}, 16934 (2015).
\bibitem{Osborn1} T. H. Osborn and A. A. Farajian, Nano Res. \textbf{7 (7)},
945 (2014).
\bibitem{Stephan} R. Stephan, M. C. Hanf, and P. Sonnet, Phys. Chem. Chem. Phys.
\textbf{17},
14495 (2015).
\bibitem{Stephan1} R. Stephan, M. C. Hanf, and P. Sonnet, J. Chem. Phys.
\textbf{143},
154706 (2015).

\bibitem{Osborn2} T. H. Osborn and A. Farajian, J. Phys. Chem. C
\textbf{116}, 22916 (2012).
\bibitem{Tritsaris} G. A. Tritsaris, E. Kaxiras, S. Meng, and E. Wang,
Nano Lett. \textbf{13}, 2258 (2013).

\bibitem{Hussain} T. Hussain, S. Chakraborty, A. D. Sarkar, B. Johansson, and 
R. 
Ahuja,
Appl. Phys. Lett. \textbf{105}, 123903 (2014).

\bibitem{Hussain1} T. Hussain, S. Chakraborty, and R. Ahuja,
Chem. Phys. Chem. Commun. \textbf{14}, 3463 (2013).

\bibitem{Song1} E. H. Song, S. H. Yoo, J. J. Kim, S. W. Lai, Q. Jiang, and S. 
O. 
Cho,
Phys. Chem. Chem. Phys. \textbf{16}, 23985 (2014).


\end{thebibliography}
\end{document}